# Efficient Graph Compression Using Huffman Coding Based Techniques

Rushabh Jitendrakumar Shah
Department of Computer Science
California State University Dominguez Hills
Carson, CA, USA
rshah13@toromail.csudh.edu

*Abstract:* **Graphs have been extensively used to represent data from various domains. In the era of Big Data, information is being generated at a fast pace, and analyzing the same is a challenge. Various methods have been proposed to speed up the analysis of the data and also mining it for information. All of this often involves using a massive array of compute nodes, and transmitting the data over the network. Of course, with the huge quantity of data, this poses a major issue to the task of gathering intelligence from data. Therefore, in order to address such issues with Big Data, using data compression techniques is a viable option. Since graphs represent most real world data, methods to compress graphs have been in the forefront of such endeavors. In this paper we propose techniques to compress graphs by finding specific patterns and replacing those with identifiers that are of variable length, an idea inspired by Huffman Coding. Specifically, given a graph $G = (V, E)$, where V is the set of vertices and E is the set of edges, and $|V| = n$, we propose methods to reduce the space requirements of the graph by compressing the adjacency representation of the same. The proposed methods show up to 80% reduction is the space required to store the graphs as compared to using the adjacency matrix. The methods can also be applied to other representations as well. The proposed techniques help solve the issues related to computing on the graphs on resources limited compute nodes, as well as reduce the latency for transfer of data over the network in case of distributed computing.**

## I. INTRODUCTION

Graphs have been studied for a long time due to their ability to represent various characteristics among data. Hence, graphs are generally used to represent data from a plethora of fields. With the advent of Big Data, graphs have become essential in representing the growing industry and academic data. After the data is gathered, there is a need to analyze the same and extract intelligence for use. For this purpose, it is essential that the graphs be computed on using various available devices.

However, due to the large size of the graphs, it is often a challenge to store the data and transfer the same over the network, specifically for using distributed or cloud computing resources. There are many applications that gain insights from analyzing graphs [9] [10]. Often, non-conventional compute units, such as GPUs are used to perform such analysis [15] [23]. However, due to the limited memory on these devices, representing the graphs using the common data structures is an issue [12] [16].

Therefore, using graph compression techniques to store graphs so that computation can be performed on the same is required. In this paper, we propose techniques to achieve lossless graph compression. Our methods are based on replacing redundant patterns with identifiers. However, since the characteristics of real world graphs indicate a power law distribution for the patterns, it is essential to not use the same size identifier for all the different patterns under consideration. Therefore, we propose using Huffman Coding based techniques, where the patterns are of variable sizes with more occurrences requiring less amount of size for the indicator. Also, the achieved compression is lossless, and can be used to represent sensitive data as well. Since the computation might be performed on resource limited devices, computing on the compressed data itself is possible [7] [13] [14].

The outline of the paper is as follows. In Section II, we discuss the previous work in this area and is summarized in the form of related work. Using Huffman Coding technique, we propose an algorithm to compress the graph, which is included in Section III. The experimental results and related analysis are provided in Section IV. Section V comprises of conclusion and future work.

## II. RELATED WORK

With a plethora of academic and business applications using graphs as the primary data structure for data storage, there have been significant previous research geared towards improving the same; this in effect increases the efficiency of the algorithms in the respective domains [17] [18] [19]. The topology of the graph is dependent on the domain of the data under consideration, and there are wide variations on the different available data.

Graph compression for real world graphs exploit the fact that these graphs are sparse and follow the power law properties. In this paper, we focus on such graphs. The previous research also focusses on such graphs and leverage the properties of the same.

The Web can be represented as a graph with the URLs as nodes and the hyperlinks between them as edges. Generally, a significant amount of compression can be achieved by modifying adjacency lists using pointers to other similar lists. Since there would be some dissimilarities, those can also be stored as an overhead in addition to the data that is shared [5]. Focusing the compression technique to work with nodes with a set of common neighbors is also another technique and is effective in compressing web graphs [1].

The concept of locality of reference can also be applied to graphs, where a set of nodes have common set of neighbors as well. This is common in graphs that represent online social networks. People tend to have a common of group of friends depending on various associations. This concept has been studied before [4].

It can be observed that proximal pages in URL lexicographical ordering often have same sets of neighbors. This lexicographical locality property allows the use of gap encodings when compressing edges. In order to further improve compression, new orderings can be developed that combine host information and Gray/lexicographic orderings [3].

If the data has a natural order, using techniques based on lexicographic ordering along with neighborhood information is effective in compressing graphs [25]; however, most real world data lack natural ordering.

Quadtree based technique to compress graph data has been used before [6] [8] [12]. However, the topological information of the graph has not been taken into consideration. Also, simple pattern matching and replacing the same with identifiers is another technique [24]. But in that case, the memory requirements for the identifiers do not represent the properties of the graph or power law appropriately. In this paper, by exploiting the distribution of the patterns in the graphs, we introduce algorithms that lead to better graph compression based on Huffman Coding principles.

### III. PROPOSED TECHNIQUES

Real world graphs are sparse and also have certain specific characteristics, like following power law distribution. Therefore, when considering the topology of the graph, there are certain patterns that can be found. In this Section, we introduce techniques to leverage these patterns to achieve graph compression. Specifically, we find the common occurring patterns of nodes and links, and replace the same with identifiers that would in essence reduce the memory requirements.

Replacing structures when some pattern is detected is a technique that has been studied before. Specifically, Quadtree is a data structure that enables such methods [8][13]. However, the identifiers that are used to replace the patterns all have the same size. In this paper, we propose an algorithm that is based on Huffman Coding principles, where we used different size identifiers for different patterns. The overall goal is to use less space to replace the most commonly occurring patterns, and more space for the less frequent ones.

We consider the adjacency matrix representation of graphs for our algorithm. Without the loss of generality, any data structure representation of graphs can be converted to an adjacency matrix representation using simple procedures. Assuming the adjacency matrix representation of graphs, we propose the usage of 32-bit patterns that would be used to identify common occurring topological structures in a graph. The 32-bit patterns can be either fixed or can be generated randomly using any known distribution. The idea of the algorithm is to find out the count of each of the patterns under consideration and then replace the patterns with certain identifiers. The patterns are divided into two groups: high occurrence patterns and low occurrence patterns.

Normally, in previous procedures, all the patterns are replaced with identifiers of the same size. However, here, we propose using different size identifiers for the high occurrence and low occurrence patterns. For each graph, we consider the 12 patterns with the highest occurrences count. From the set of 12, we choose the top 4 counts to belong to the high occurrence group, and the rest 8 belong to the low occurrence group. Since there are only 4 patterns in the high occurrence group, each can be uniquely identified using 2-bit identifiers; for the low occurrence group, the 8 different patterns can be uniquely identified using 3-bit identifiers. For non-matches, the raw data would be stored. The first bit of any sequence would indicate whether the following bits belong to an identifier for a pattern or raw data. A 0 indicates the following 32-bits are raw data, and a leading 1 indicates the following are identifiers. Now, since there are 2 groups for the patterns, a 0 indicates a high occurrence patterns and a 1 indicates a low occurrence pattern. For the high occurrence patterns, the identifier is a leading 0 followed by 2-bits to identify the pattern, and for the low occurrence patterns, the identifier is a leading 1 followed by 3-bits to identify the pattern. Hence, in total there are 3 types of sequences that replaces the chunks of 32 bits from the adjacency matrix. In case of pattern matching with high occurrence patterns, the 32 bits are replaced by 4 bits, in case of low occurrence patterns the 32 bits are replaced by 5 bits. This method helps reduce the memory requirement for the patterns matched. However, for all the 32-bit chunks from the adjacency matrix that does not match any of the patterns under consideration, the 32-bit chunks are replaced by 33 bits. In addition to the raw 32 bits that are stored without any modification from the adjacency matrix, there is a 1 bit overhead to indicate that the data is not an identifier. This method of compressing the adjacency matrix information is given in Algorithm 1.

Algorithm 1: Matching patterns in adjacency matrix
Input: Adjacency matrix A[ ] for G = (V, E), Pattern Set {P}
Output: Compressed matrix $A_c$[ ] for G = (V, E)
For each row in A[ ]
    Divide row into chunks, Size(chunk)=Size(patterns)
    For each chunk in the row
        If matchPattern(chunk, $P_i$), $P_i \in$ {P}.
            Insert 1
            If High Occurrence Pattern
                Insert 2-bit identifier
            Else
                Insert 3-bit identifier
        Else
            Insert 0
            Insert Chunk

Algorithm 1 splits each of the rows of the adjacency matrix of the graph under consideration into chunks of sizes equal to those of the patterns. The chunks are then matched with patterns in the set of patterns given by {P}. As discussed earlier, the patterns in {P} can be either fixed or variable. Each successful match results in decrease of the memory requirement for the graph, and each non-match needs a penalty and adds overhead.

Consider the number of patterns to be matched is $n$ and the size of each of the patterns being matched is $s$. The size of the adjacency matrix for given graph G with $v$ nodes would be $v*v$ bits. In case of using same number of bits for each of the identifiers, the size of each identifier is $\log_2 n$. Since the size of the chunks to be matched is the same as the size of the patterns, the total number of chunks for the given graph G is $(v/s)*v$.

Therefore, in the ideal case, where all the chunks are matched to respective patterns, the size of the compressed graph $G_c$ is given by

$$G_c = (1 + \log_2 n) * ((v/s)*v) \quad (1)$$

The additional 1 bit added to size of the identifier is for the indicator to determine the data is not raw data, rather it belongs to a pattern.
In case of where there are $m$ matches, the size of the compressed graph is given by

$$G_c = (1 + \log_2 n) * m + (1+s)*((v/s)*v - m) \quad (2)$$

In our proposed technique, we use two different groups of patterns, high occurrence and low occurrence. Now, considering the total number of patterns to be matched is still $n$ and the size of the patterns being matched is still $s$, we divide the number of patterns $n$ into $h$ and $l$, which indicates the number of high occurrences and low occurrences.

$$n = h + l \quad (3)$$

Considering there are still $m$ matches and there are $m_h$ and $m_l$ matches for the high and low occurrences respectively, the size of the compressed graph for our algorithm is given by,

$$G_c = (1 + \log_2 h)*m_h + (1 + \log_2 l)*m_l + (1+s)*((v/s)*v - m) \quad (4)$$

## IV. EXPERIMENTAL RESULTS

In this Section, we present the results of the proposed algorithm on sample graphs. Different size graphs are considered, that represent the properties of real world graphs, and the observations are discussed.

For the experiments, we consider 4 different graphs with 1024, 2048, 4096 and 8192 nodes. Real world graphs have huge number of nodes in them; however, analysis has shown the number of nodes under consideration for most calculations are based on the size of the largest connected components that exist in these graphs, and those are similar to the size of graphs we have considered [10] [11]. Therefore, our results and observations are significant for real world data [20] [21] [22].

The experimental results for the number of patterns found for the graphs under consideration is in Fig. 1. Since the size of the patterns are fixed at 32 bits, the maximum number of possible pattern matches is given by $(v*v)/32$. For graphs with 1024, 2048, 4096 and 8192 nodes, the values are 32768, 131072, 524288 and 2097152 respectively. In Fig. 1, for each of the graphs under consideration, plots are provided for the maximum possible patterns, total patterns matched, total number of high occurrences patterns found and total number of low occurrences patterns found.

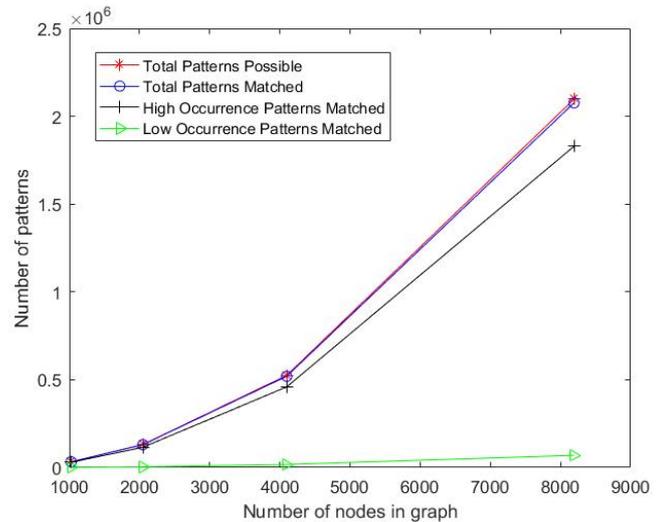

Fig. 1: Pattern counts

It can be observed from the plots in Fig. 1, that the number of patterns of high occurrence is significantly high. This in turn increases the efficiency of the compression algorithm.

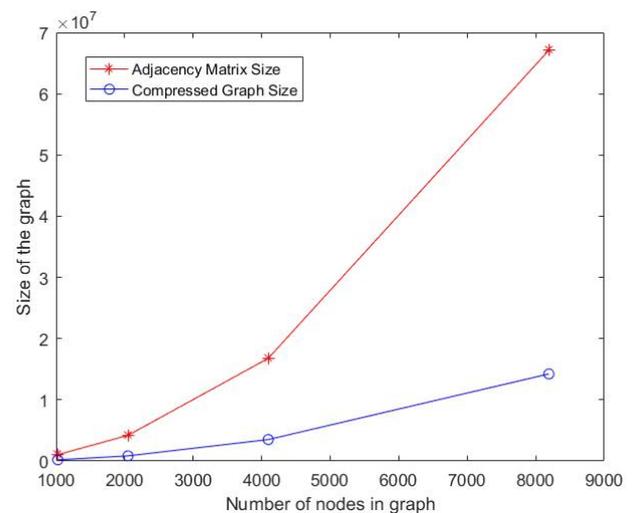

Fig. 2: Compression achieved

The compression algorithm is also implemented using the graphs as described. The results are plotted in Fig. 2. As it can be observed, for all the different graphs, the compression achieved is almost 80% as compared to the adjacency matrix representation.

## V. CONCLUSION

Graphs represent one of the most commonly used data structures to characterize real world information. In the era of Big Data there is a need to represent data in the most efficient manner in order for efficient processing and reduced data transfer overhead. In this paper, we propose graph compression techniques based on pattern matching and using Huffman coding principles to reduce the memory requirement for adjacency matrix representation of graphs. In particular, we split patterns into high and low occurrences and use lower number of bits for the high occurrence patterns and vice-versa. The simulation results show that using the proposed algorithm, we can achieve up to 80% compression as compared to the adjacency matrix representation of graphs. The proposed algorithm works primarily with adjacency matrix for graphs, however the proposed techniques would work with other techniques such as Quadtree compression. Our future work would involve generation of patterns that can be easily utilized on the data based on the domain.

REFERENCES

[1] Micah Adler and Michael Mitzenmacher. Towards Compressing Web Graphs. In Proceedings of the Data Compression Conference, pages 203–212, Washington, DC, USA, 2001.

[2] David A. Bader and Kamesh Madduri. GTgraph: A Synthetic Graph Generator Suite, 2006.

[3] P. Boldi and S. Vigna. The Webgraph Framework I: Compression Techniques. In Proceedings of the 13th International Conference on World Wide Web, WWW '04, pages 595–602, New York, NY, USA, 2004. ACM.

[4] J. Leskovec, K. Lang, A. Dasgupta, and M. Mahoney. Community Structure in Large Networks: Natural Cluster Sizes and the Absence of Large Well-Defined Clusters. Internet Mathematics, Vol. 6(1):29–123, 2009.

[5] Keith H. Randall, Raymie Stata, Janet L. Wiener, and Rajiv G. Wickremesinghe. The Link Database: Fast Access to Graphs of the Web. In Proceedings of the Data Compression Conference, DCC '02, pages 122–131, Washington, DC, USA, 2002. IEEE Computer Society.

[6] Hanan Samet. Using quadtrees to represent spatial data. In Herbert Freeman and GoffredoG. Pieroni, editors, Computer Architectures for Spatially Distributed Data, volume 18 of NATO ASI Series, pages 229– 247. Springer Berlin Heidelberg, 1985.

[7] A. Chatterjee, A. Aceves, R. Dungca, H. Flores, and K. Giddens. Classification of wearable computing: A survey of electronic assistive technology and future design. In Research in Computational Intelligence and Communication Networks (ICRCICN), pages 22–27. IEEE, 2016.

[8] A. Chatterjee, M. Levan, C. Lanham, M. Zerrudo, M. Nelson, and S. Radhakrishnan. Exploiting topological structures for graph compression based on quadtrees. In Research in Computational Intelligence and Communication Networks (ICRCICN), 2016 Second International Conference on, pages 192–197. IEEE, 2016.

[9] A. Chatterjee, S. Radhakrishnan, and J. K. Antonio. Data Structures and Algorithms for Counting Problems on Graphs using GPU. International Journal of Networking and Computing (IJNC), Vol. 3(2):264–288, 2013.

[10] A. Chatterjee, S. Radhakrishnan, and J. K. Antonio. On Analyzing Large Graphs Using GPUs. In Parallel and Distributed Processing Symposium Workshops & PhD Forum (IPDPSW), 2013 IEEE 27th International, pages 751–760. IEEE, 2013.

[11] A. Chatterjee, S. Radhakrishnan, and John K. Antonio. Counting Problems on Graphs: GPU Storage and Parallel Computing Techniques. In Parallel and Distributed Processing Symposium Workshops & PhD Forum, 2012 IEEE 26th International, pages 804–812. IEEE, 2012.

[12] M. Nelson, S. Radhakrishnan, A. Chatterjee, and C. N. Sekharan. On compressing massive streaming graphs with Quadtrees. In Big Data, 2015 IEEE International Conference on, pages 2409–2417, 2015.

[13] M. Nelson, S. Radhakrishnan, A. Chatterjee and C. N. Sekharan, "Queryable compression on streaming social networks," 2017 IEEE International Conference on Big Data (Big Data), Boston, MA, 2017, pp. 988-993.

[14] A. Chatterjee, M. Levan, C. Lanham and M. Zerrudo, "Job scheduling in cloud datacenters using enhanced particle swarm optimization," 2017 2nd International Conference for Convergence in Technology (I2CT), Mumbai, 2017, pp. 895-900.

[15] Khondker S. Hasan, Amlan Chatterjee, Sridhar Radhakrishnan, and John K. Antonio, "Performance Prediction Model and Analysis for Compute-Intensive Tasks on GPUs", Network and Parallel Computing 2014, pp. 612-617

[16] A. Chatterjee, "Parallel Algorithms for Counting Problems on Graphs Using Graphics Processing Units", Ph.D. Dissertation, University of Oklahoma, 2014

[17] A. Chatterjee, H. Flores, S. Sen, K. S. Hasan and A. Mani, "Distributed location detection algorithms using IoT for commercial aviation," 2017 Third International Conference on Research in Computational Intelligence and Communication Networks (ICRCICN), Kolkata, 2017, pp. 126-131.

[18] A. Chatterjee, H. Flores, B. Tang, A. Mani and K. S. Hasan, "Efficient clear air turbulence avoidance algorithms using IoT for commercial aviation," 2017 Recent Developments in Control, Automation & Power Engineering (RDCAPE), NOIDA, India, 2017, pp. 28-33.

[19] A. Chatterjee, S. Radhakrishnan, C.N. Sekharan, " Connecting the dots: Triangle completion and related problems on large data sets using GPUs," 2014 IEEE International Conference on Big Data (Big Data), Washington, DC, 2014, pp. 1-8.

[20] A. Chatterjee, J. Chen, M. Perez, E. Tapia and J. Tsan, "Energy efficient framework for health monitoring using mobile systems," 2017 2nd International Conference for Convergence in Technology (I2CT), Mumbai, 2017, pp. 1116-1120.

[21] Ishita Das, Santanu Roy, Amlan Chatterjee, Soumya Sen, "A Data Warehouse Based Schema Design on Decision Making in Loan Disbursement for Indian Advance Sector", IEMIS 2018, Springer.

[22] Sinthia Roy, Arijit Banerjee, Partha Ghosh, Amlan Chatterjee, Soumya Sen, "Intelligent Web Service Searching Using Inverted Index", Springer, ICCAIAIT 2018.

[23] Khondker S. Hasan, Mohammad Rob, Amlan Chatterjee, "Correlation based Empirical Model for Estimating CPU Availability for Multi-Core Processor in a Computer Grid, PDPTA 2018.

[24] Shah, Rushabh Jitendrakumar. "Graph Compression Using Pattern Matching Techniques." arXiv preprint arXiv:1806.01504 (2018).

[25] Flavio Chierichetti, Ravi Kumar, Silvio Lattanzi, Michael Mitzenmacher, Alessandro Panconesi, and Prabhakar Raghavan. On Compressing Social Networks. In Proceedings of the 15th ACM SIGKDD International Conference on Knowledge Discovery and Data Mining, KDD '09, pages 219–228, New York, NY, USA, 2009. ACM.